\documentclass{ws-procs9x6}

\begin{document}

\title{MEASUREMENTS OF UNPOLARIZED AZIMUTHAL ASYMMETRIES AT COMPASS}

\author{W. K\"afer on behalf of the COMPASS collaboration}

\address{Physikalisches Institut, Albert-Ludwigs-University Freiburg,\\
79104 Freiburg, Germany\\
E-mail: wolfgang.kaefer@ik.fzk.de\\
}

\begin{abstract}
Azimuthal Asymmetries in unpolarized SIDIS can be used to probe the transverse
momentum of quarks inside the nucleon. Furthermore, they give access to
the so-far unmeasured Boer-Mulders function. We report on the first
measurement of azimuthal asymmetries of the SIDIS cross section from scattering of muons off a deuteron target. 
\end{abstract}


\bodymatter

\section{Introduction and Theoretical Motivation}\label{sec:Theory}

\begin{figure} [hbpt]
\centering
\psfig{file=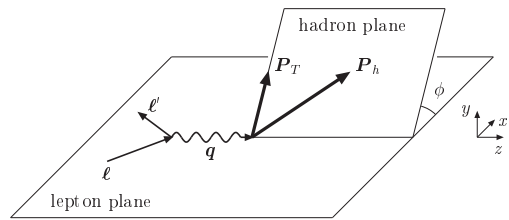, width=0.73\textwidth}
\caption{\label{Fig:DEFPHI}Definition of the hadron azimuthal angle ${\phi_h}$}
\end{figure}

 Semi Inclusive Deep Inelastic Scattering (SIDIS) reactions are an important
 tool to probe the structure of the nucleon. Of particular interest in
 unpolarized SIDIS is a possible dependence of the cross section on the
 azimuthal angle of the hadron as defined in Fig.~\ref{Fig:DEFPHI}.  

There are several effects which contribute to a $\cos\phi_h$ and $\cos2\phi_h$
dependence of the cross section. The amplitudes of these modulations depend on the kinematic variables
relevant for the SIDIS process, namely the Bjorken scaling variable $x$, the
relative energy loss of the muon $y$, the negative 4-momentum of the virtual
photon squared $Q^2$, the energy fraction of the hadron $z$ and the transverse
momentum of the hadron $P_t^h$.
 Furthermore, $k_t$ is the transverse momentum of the quark with respect to the
nucleon, and $p_t$ the transverse momentum of the hadron with respect to the direction
of the fragmenting quark.

The first azimuthal dependence discussed here is due to non-zero $k_t$:
As Cahn pointed out in Ref.~\refcite{CAHN}, an azimuthal modulation of the cross
 section is expected in one-photon exchange approximation, when the transverse
 momentum of the quark is taken into account. The Cahn effect contributes to a
 possible $\cos\phi_h$ and $\cos2\phi_h$ asymmetry of the SIDIS cross
 section. It is kinematically suppressed by $\frac{k_t}{Q}$ for
 $\cos(\phi_q)$ and $\left( \frac{k_t}{Q} \right)^2$ for the $\cos2\phi_q$
 term. When going from the quark to hadron level, the unpolarized parton distribution functions (PDFs) $f_q(x, k_t)$ and fragmentation functions (FFs) $D_q^h(z, p_t)$ need to be taken into account. Assuming Gaussian
 distributions for the transverse momentum dependence of $f_q(x, k_t)$ and $D_q^h(z, p_t)$ and introducing $D_{\cos\phi_h}(y) = \frac{ (2-y)\sqrt{1-y}}{1 + (1+y)^2}$, the $\cos\phi_h$ term can be written as
\begin{equation}
    \frac{\textrm{d}^5\sigma} {\textrm{d}x\, \textrm{d}y\, \textrm{d}z\, \textrm{d}{P_t^h}^2\, d\phi_h} \propto \sum_q f_q(x) D_q^h(z) \left[ 1-4 D_{\cos\phi_h}(y) \frac{\langle k_t^2 \rangle z P_t^h} {Q \langle \left( P_t^h\right) ^2 \rangle } \cos\phi_h \right] \;.
\end{equation}
The $\cos\phi_h$ modulation allows the determination of $\langle k_t^2 \rangle$.
 This has been done e.g. in Ref.~\refcite{ANSELMINO} using results from previous
 experiments, arriving at $\langle k_t^2 \rangle \approx 0.25\mbox{ GeV$^2$/$c^2$}$.

The second contribution to azimuthal asymmetries comes from the Boer-Mulders PDF~\cite{PhysRevD.57.5780} $h_1^\perp(x, k_t)$, convoluted with the Collins FF $H_1^\perp(x, p_t)$.
A model calculation~\cite{Barone:2008tn} shows that the Boer-Mulders
contribution to the $\cos2\phi_h $ modulation might be of similar magnitude
than the contribution from the Cahn effect. In particular, a possible difference between positive
and negative hadron asymmetries could be explained by the Boer-Mulders mechanism. It may also contribute to the $\cos\phi_h$ asymmetry~\cite{PhysRevD.57.5780},
but the size of this effect is so far unknown and no predictions exist.

Perturbative QCD (pQCD) introduces a third $\phi_h$ dependence at order
$\alpha_s$. However, QCD effects are only important at $P_t^h >$ 1
GeV/c. Therefore they are expected to be small for COMPASS kinematics, where
most of the statistics is at low transverse momentum. The perturbative
contribution has been given in Refs.~\refcite{GEORGI, QCD_COS2} at ${\cal O}
(\alpha_s^1 )$. Recently, higher order contributions were calculated in Ref.~\refcite{QCD2}.

Since the muon beam used at the COMPASS experiment is naturally polarized, an additional modulation of the cross section is expected. In contrast to the Cahn, Boer - Mulders and pQCD contributions, the cross section for the beam asymmetry depends on $\sin \phi_h$~\cite{Kotzinian}. This effect has recently been measured by the CLAS collaboration~\cite{CLAS} on a proton target. 
 
Radiative corrections and possible higher twist terms may contribute to
the modulations, but these effects are considered to be small and therefore will not be discussed here.

 The overall cross section for SIDIS on an unpolarized target is thus of the form:
\begin{equation}
 \frac {\textrm{d}\sigma} {\textrm{d}\phi_h} \propto 1 + A_{\cos\phi_h } \cos \phi_h
 + A_{\cos2\phi_h} \cos2\phi_h + A_{\sin\phi_h} \sin\phi_h \;.
 \end{equation}
The amplitudes of these three modulations have been determined from data taken
with the COMPASS experiment with a deuteron target.

For a proton target, three experiments have published results on azimuthal asymmetries in different kinematic regions: EMC~\cite{EMC_1, EMC_2}, the E665
collaboration~\cite{PhysRevD.48.5057} and the ZEUS
experiment~\cite{ZEUS}\@. 

 \section{The COMPASS Experiment}\label{sec:COMPASS}
The COMPASS experiment~\cite{COMPASS_spec} is a fixed target experiment at CERN. It features a 160
GeV/c $ \mu^+$-beam, with a natural polarization of -80\% and a
polarized target, which consists of two cells. From 2002-2006, data was taken with a
polarized ${}^6\mathrm{Li} \mathrm{D}$ target, while in 2007 a hydrogen target was used.
The target can be polarized either in the longitudinal or the transverse direction with respect to the beam direction. These
two configurations will be called CL and CT in the following.

 The different magnetic fields needed to maintain the
target polarization (a solenoid field for CL and a
dipole field for CT), require changes of the magnetic
configuration of the experiment. In particular, the beam line settings differ, since an additional kick is needed for the CT setup to compensate the additional dipole field. On the other hand, for CL, there is a
strong interference between the target magnetic field and the magnetic field
of the first spectrometer magnet, which is not present for
CT.

\section{Asymmetry Extraction}\label{sec:METHOD}

The data sample used for this analysis was taken in the year 2004. Data
samples in both target configurations (CL and CT) have been used in order
to get a better estimate of the systematic error generated by experimental conditions.

 The final sample consists of about 5 million positive and 4 million negative hadrons in the
 kinematic range $Q^2>1 \mbox{ GeV$^2$/$c^2$}$,$ 0.1<y<0.9$, $0.2<z<0.85$ and 
  $0.1 \mbox{ GeV/c}< P_t^h < 1.5 \mbox{ GeV/c}$ and contains data from both
 target configurations in roughly equal parts.
 The target polarization was canceled by event weighting, taking into
 account the fluxes and average polarizations: 
\begin{equation}
\label{EQ:WEIGHTING}
 N_{\textrm{unpol}}= P_2 N_1 + F P_1 N_2\;.
\end{equation}
where $P_i$ indicates the absolute value of the polarizations for the two possible spin
configurations ($\leftarrow, \rightarrow$ for CL and $\uparrow, \downarrow$ for CT), $N_i$ the event number for the respective polarization and $F=\frac{F_1}{F_2}$ the corresponding ratio of the respective muon fluxes.

\begin{figure} [hbpt]
\centering
\psfig{file=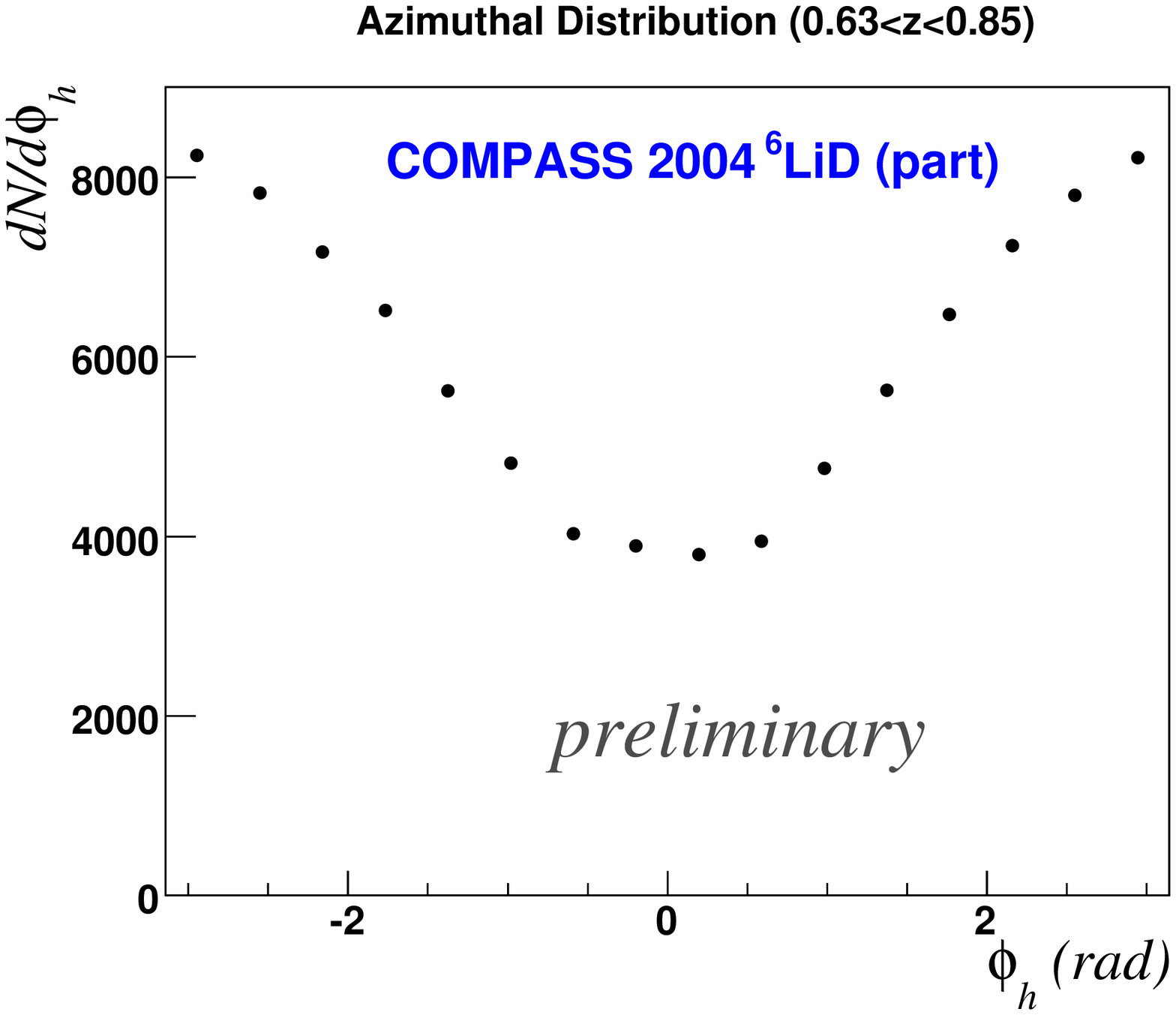, width=0.45\textwidth, trim= 0 0 00 50,clip}
\psfig{file=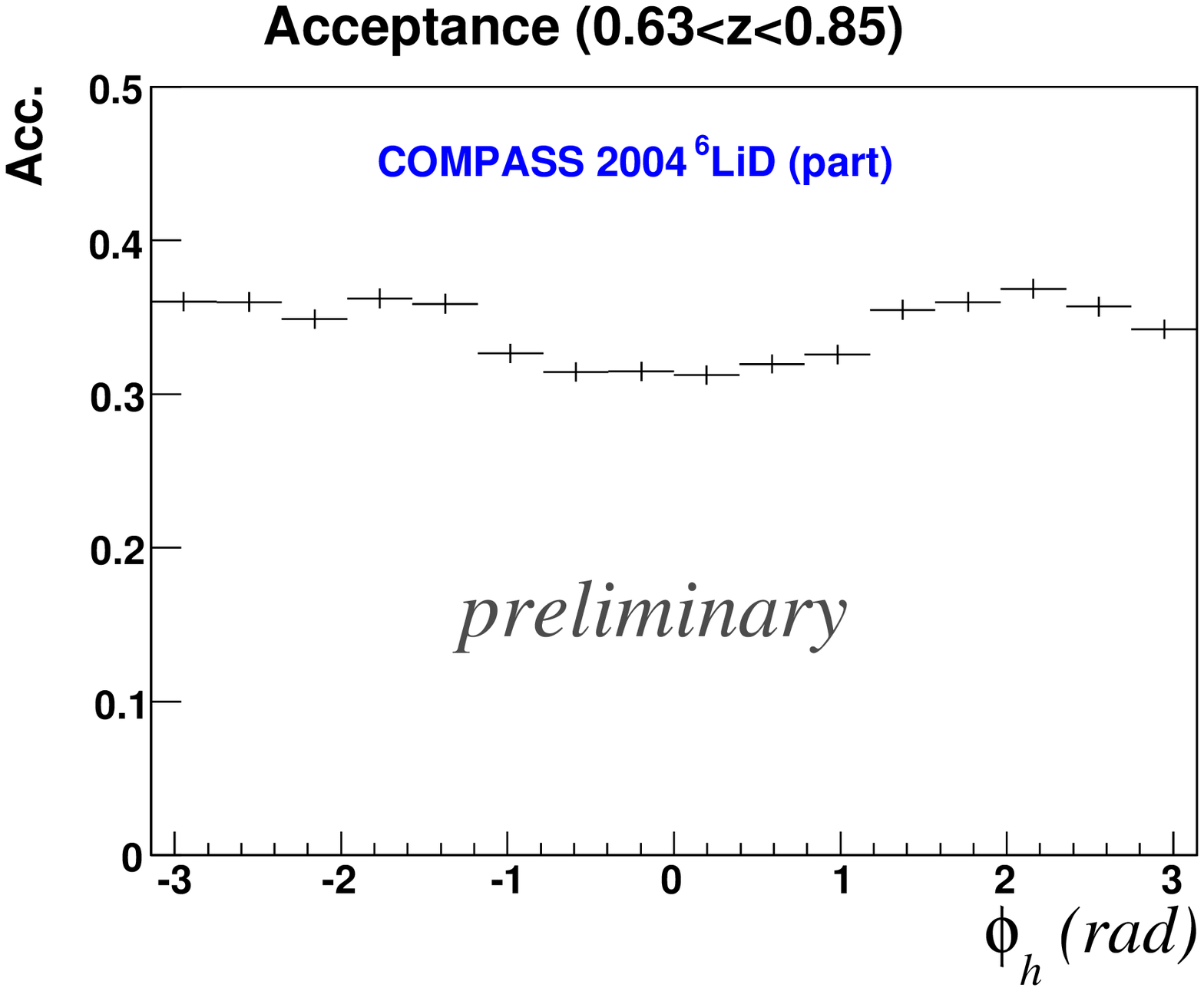, width=0.45\textwidth, trim= 0 0 00 45,clip}
\caption{measured $\phi_h$ distribution before correction and acceptance in the CT case for $0.63 < z < 0.85$. }
\label{Fig:Acc}
\end{figure}

In order to correct for event losses caused by the non-uniform acceptance of
the COMPASS spectrometer, full MC simulations have been performed in both
CL and CT setup. In each case, the events were generated
with LEPTO, transported through the COMPASS detector simulation program
COMGEANT and the reconstruction software CORAL. From these
MC samples, the acceptance of the COMPASS spectrometer $A(\phi_h)$ and the corrected count rates $N_{\textrm{corr}} (\phi_h) $ can be determined:
 \begin {equation}
A(\phi_h)= \frac{N_{\textrm {rec}}(\phi_h)}{ N_{\textrm{gen}}(\phi_h) } \qquad \qquad
 N_{\textrm{corr}}(\phi_h) = \frac{N_{\textrm{unpol}} (\phi_h) } { A(\phi_h)}\;.
\end{equation}
The acceptance correction has been done in bins of $x, z$, and $P_t^h$
 separately, with the other two variables always integrated out. Fig.~\ref{Fig:Acc} shows a typical example of measured $\phi_h$ distribution
 and the corresponding acceptance.
The corrected count rates are then fitted with a four-parameter fit containing the average count rate $N_0$ and the three amplitudes:
\begin{equation}
 N_\textrm{corr} = N_0 ( 1 + A_{\sin\phi_h} \sin \phi_h + A_{\cos\phi_h}\cos \phi_h + A_{\cos2\phi_h}\cos
 2 \phi_h )\;.
\end{equation}

\section{Results}\label{sec:RESULTS}
\begin{figure} [hbpt]
\centering
\psfig{file=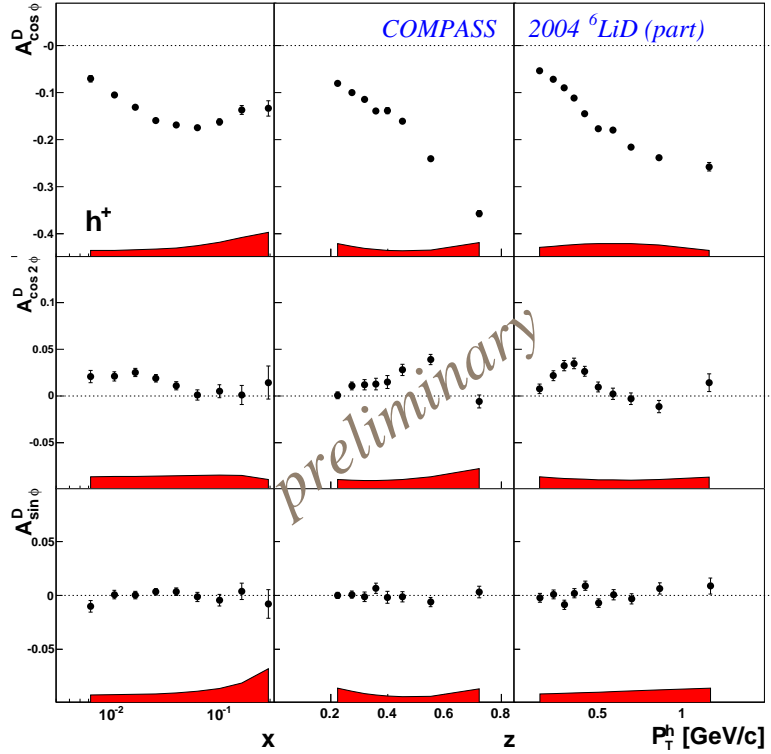, width=0.9\textwidth}
\caption{Results for positive hadrons for $A_{\cos \phi_h}$, $A_{\cos 2\phi_h}$ and $A_{\sin \phi_h}$ in dependence of the kinematic variables $x, z, P_t^h$. The error bars correspond to the statistical errors, while the error bands at the bottom indicate systematic errors.}
\label{Fig:RESULTSPOS}
\end{figure}

\begin{figure}[hbpt]
\centering
\psfig{file=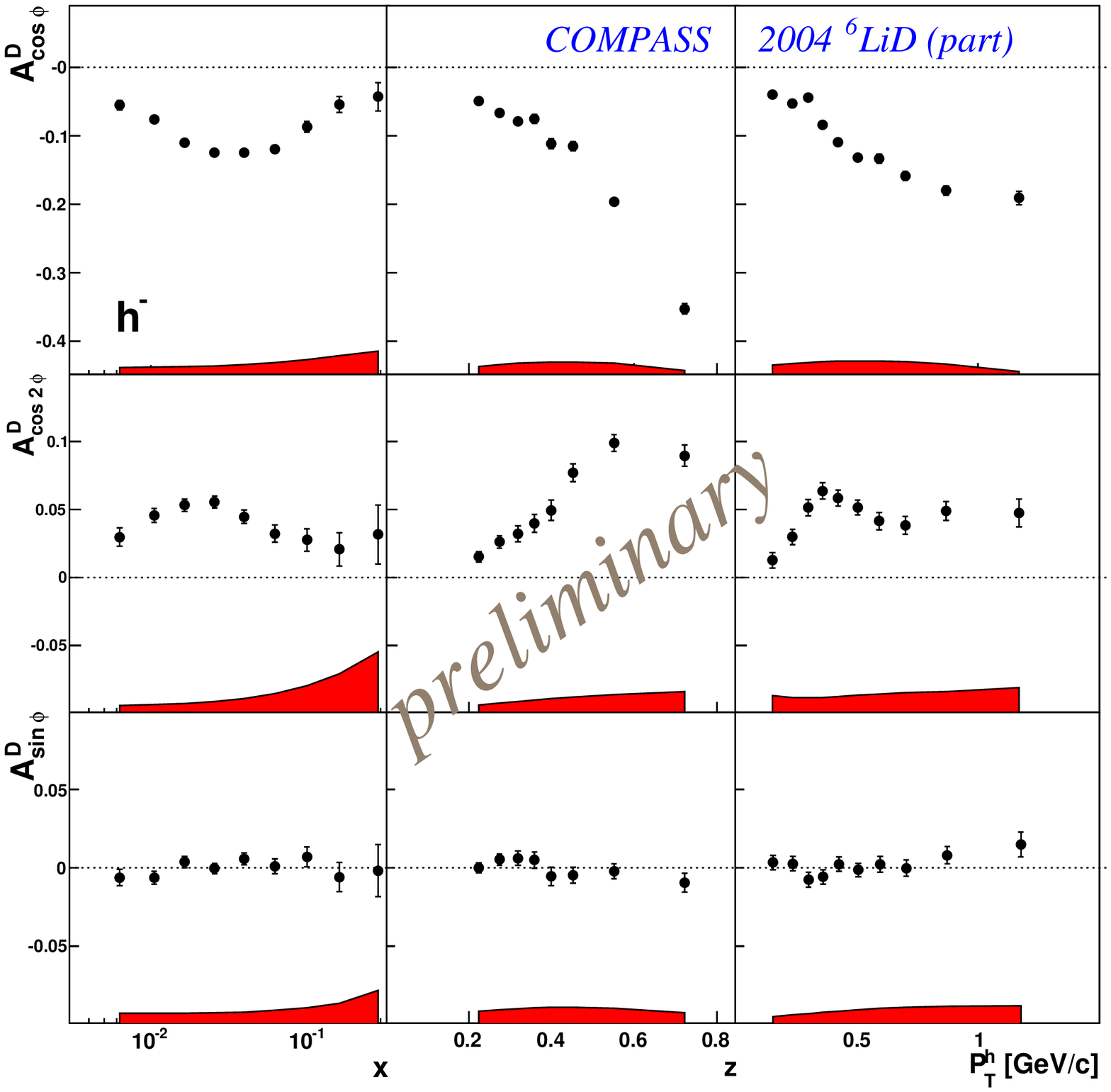, width=0.9\textwidth}
\caption{Results for negative hadrons for $A_{\cos \phi_h}$, $A_{\cos 2\phi_h}$ and $A_{\sin \phi_h}$ in dependence of the kinematic variables $x, z, P_t^h$. The meaning of the error bars and bands is the same as in the previous figure.}
\label{Fig:RESULTSNEG}
\end{figure}

The amplitudes $ A_{\cos\phi_h }, A_{\cos2\phi_h }$ and $ A_{\sin\phi_h }$ of the
three modulations have been determined in dependence on $x, z$ and $
P_t^h$ for positive and negative hadrons seperately. Fig.~\ref{Fig:RESULTSPOS} shows the results for the three modulations
for positive hadrons. A large $\cos\phi_h$ amplitude of up to 40\% and a
$\cos2\phi_h$ amplitude of up to 5\% is seen. The value for $A_{\sin\phi_h}$ are
consistent with zero. Also for the negative hadrons, shown in
Fig.~\ref{Fig:RESULTSNEG}, the $\sin\phi_h$ amplitude is consistent with
zero, while the two cosine amplitudes show similar trends compared to the
positive hadrons. However, the magnitude of the cosine modulations differs significantly
for positive and negative hadrons.

\begin{figure} [hbpt]
\centering
\psfig{file=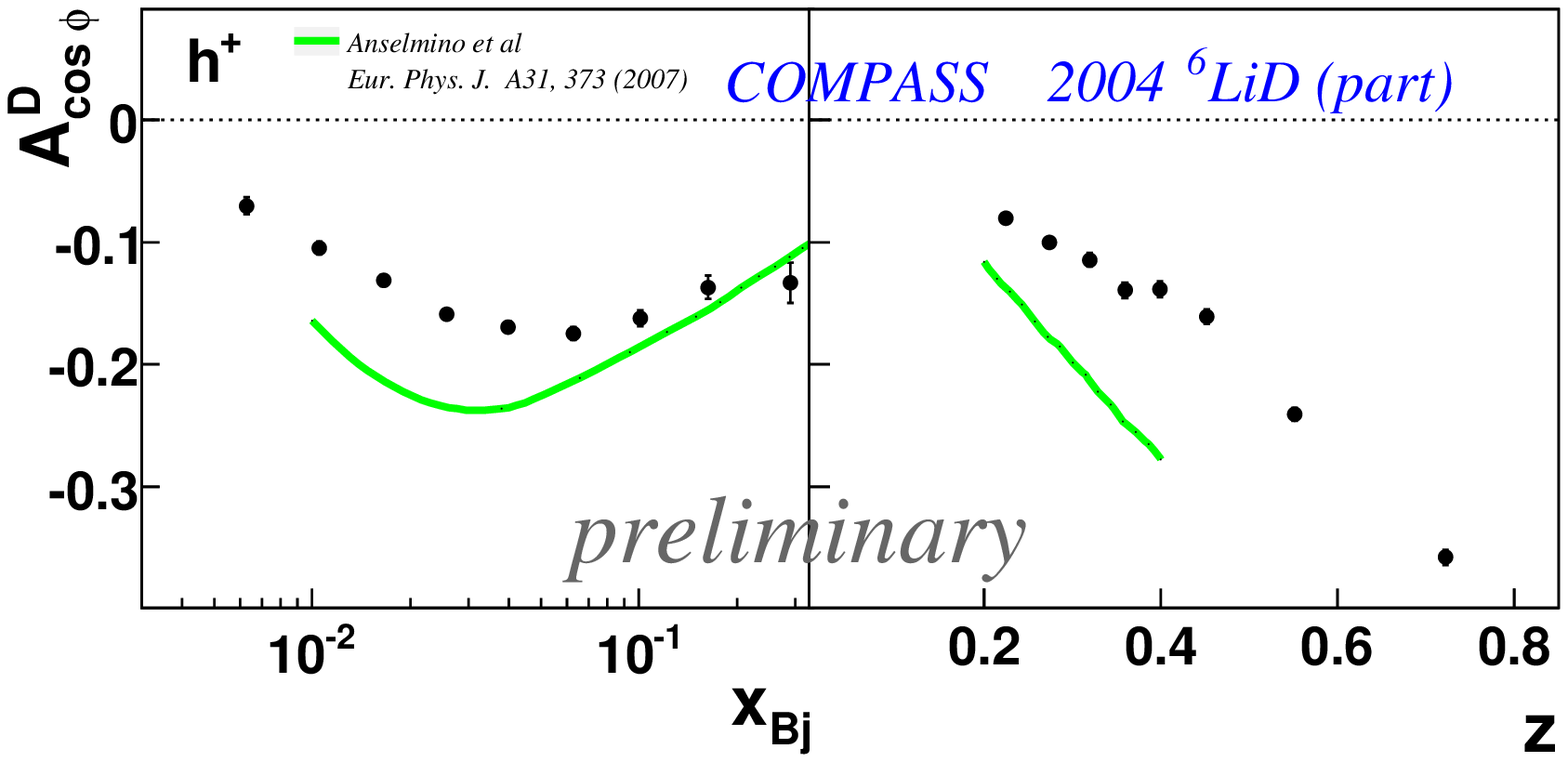, width=0.9\textwidth, trim= 0 0 00 0,clip}
\caption{Comparison of $A_{\cos \phi_h}$ for positive hadrons with model calculation from Ref.~\protect\refcite{Anselmino:2006rv}, which takes into account Cahn and perturbative QCD effects. Only statistical errors are shown.}
\label{Fig:COMPCOS}
\end{figure}

\begin{figure} [hbpt]
\centering
\psfig{file=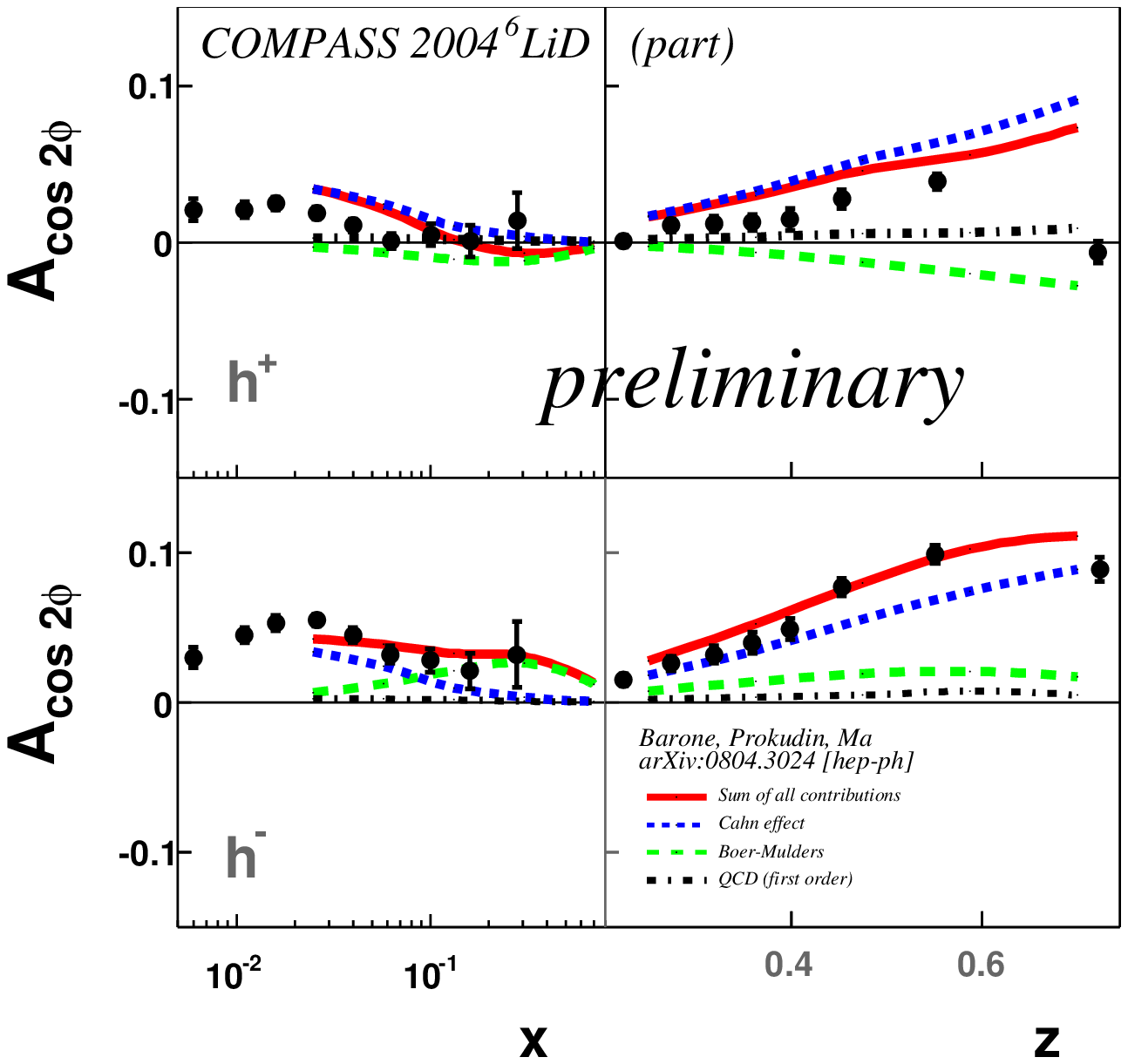, width=0.9\textwidth,  trim= 0 0 75 0,clip}
\caption{Comparison of $A_{\cos 2\phi_h}$ with predictions for COMPASS kinematics from Ref.~\protect\refcite{Barone:2008tn}. The calculation takes into account the Cahn effect(dashed curve), perturbative QCD (dashed dotted) and Boer-Mulders (dotted curve). The continuous line describes the sum of these three. Again, only statistical errors are shown.}
\label{Fig:COMPCOS2}
\end{figure}

 Fig.~\ref{Fig:COMPCOS} shows a recent prediction~\cite{Anselmino:2006rv} for a deuteron target and the COMPASS
kinematics, compared to the data. Only Cahn and perturbative QCD
contributions have been considered. Calculation and data show similar behaviour in the region of large $x$, while
there is disagreement in the region of small $x$. The strong disagreement in the low $x$ region leads to a scale difference in $z$, although the slope, which is mainly due to the Cahn Effect, is again similar.
A model-calculation~\cite{Barone:2008tn} for the $\cos2\phi_h$ amplitude is
compared to the COMPASS results in Fig.~\ref{Fig:COMPCOS2}. Here also the
Boer-Mulders mechanism has been included. Since $h_1^\perp(x, k_t)$ is presently not constrained by experimental data, it has been assumed to be proportional to the better known
Sivers function. This assumption has e.g. been motivated in Ref.~\refcite{Burkardt:2005hp}.

\section{Systematic tests}  {\label{sec:Syst}
Several checks have been performed to determine the systematic
error of the measurement. It turns out that the systematic error is due to two sources: the difference between the results in the setups CL and CT and the variation of the acceptance for different generated kinematic distributions. These differences are used as an estimate for the quality of the acceptance correction. To evaluate the second contribution, for each setup, CL and CT, two MC simulations with different LEPTO settings and thus different generated kinematic distributions have been performed. Several additional tests, such as splitting the data sample according to the event topology, target polarization and time of the measurement give no significant contribution to the systematic error.

\section{Summary and Outlook}\label{sec:SUMMARY}
First results on unpolarized azimuthal asymmetries from COMPASS
have been presented, extending the investigated kinematic region to low $x$.
The data can be used to better determine the value of $\langle k_t^2
\rangle$. Also the differences between positive and negative hadrons allow to gain knowledge about $h_1^\perp(x, k_t)$, which can be further deepened with the data taken in 2007 with a $\textrm{NH}_3$ target. 

\bibliographystyle{ws-procs9x6}
\bibliography{Unpolarized_Asym_COMPASS}

\end{document}